\documentclass[aps,prd,twocolumn,showpacs,eqsecnum,nofootinbib]{revtex4}

\usepackage{epsfig}
\usepackage{graphicx}
\usepackage{dcolumn}
\usepackage{amsmath}
\usepackage{enumerate}
\usepackage{epstopdf} 

\newcommand{\tfo}{T_{\rm f.o.}}
\newcommand{\trh}{T_{\rm RH}}
\newcommand{\sigmav}{{\langle\sigma v\rangle}}
\renewcommand{\Re}{\mathop{\rm Re}}

\begin{document}

\title{%
\vskip-6pt \hfill {\rm\normalsize UCLA/06/TEP/07} \\
\vskip-12pt~\\
Neutralino with the Right Cold Dark Matter Abundance\\
in (Almost) Any Supersymmetric Model }

\author{
\mbox{Graciela Gelmini$^{1}$} and
\mbox{Paolo Gondolo$^{2}$}}

\affiliation{
\mbox{$^1$  Department of Physics and Astronomy, UCLA,
 475 Portola Plaza, Los Angeles, CA 90095, USA}
\mbox{$^2$ Department of Physics, University of Utah,
   115 S 1400 E \# 201, Salt Lake City, UT 84112, USA }
\\
{\tt gelmini@physics.ucla.edu},
{\tt paolo@physics.utah.edu}}


\vspace{6mm}
\renewcommand{\thefootnote}{\arabic{footnote}}
\setcounter{footnote}{0}
\setcounter{section}{1}
\setcounter{equation}{0}
\renewcommand{\theequation}{\arabic{equation}}

\begin{abstract} \noindent
We consider non-standard cosmological models in which the late 
decay of a scalar field $\phi$ reheats the Universe to a low reheating temperature,
between 5 MeV and the standard freeze-out temperature of neutralinos
of mass $m_{\chi}$. We point out that  in these models all 
 neutralinos  with  standard density
  $\Omega_{\rm std} \gtrsim 10^{-5} (100 {\rm GeV}/m_\chi)$ can have the density
of cold dark matter, provided the 
  right combination of the following  two parameters can be achieved in the high energy theory:
 the reheating temperature, and the ratio of the
 number of neutralinos produced per $\phi$ decay over the $\phi$ field mass.
 We present the ranges of these parameters where  a combination of thermal
and non-thermal neutralino production leads to the desired density,
as functions of $\Omega_{\rm std}$ and  $m_{\chi}$.
\end{abstract}

\pacs{14.60.St, 98.80.Cq}

\maketitle 

In supersymmetric models, the lightest supersymmetric particle (LSP),
usually a neutralino $\chi$, is a good cold dark matter candidate. The
cosmological density of the neutralino is a function of the
supersymmetric model parameters, and it has been computed
theoretically to high precision. Requiring the LSP to have the
measured dark matter density constrains the models considerably to
very narrow regions in parameter space.

The standard computation of the relic density relies on the
assumptions that the entropy of matter and radiation is conserved,
that neutralinos are produced thermally and were in thermal and
chemical equilibrium before they decoupled.  The decoupling, or
freeze-out, temperature, i.e.\ the temperature after which their
number practically does not change, is $\tfo \simeq m_{\chi}/20$,
where $m_\chi$ is the neutralino mass.  The standard neutralino relic density
$\Omega_{\rm std}$ obtained in this way can be larger or smaller than
the measured density of cold dark matter $\Omega_{\rm cdm} =
0.113\pm0.009 h^{-2}$ \cite{wmap}.

However, there are non-standard cosmological models in which the
assumptions mentioned above do not hold. These include models with
moduli decay \cite{Moroi-Randall}, Q-ball decay \cite{Fujii}, and
thermal inflation \cite{Lazarides}. In all of these models there is a
late episode of entropy production and  non-thermal
production of the LSP in particle decays is possible.

We concentrate on cosmological models in which the late decay of a
scalar field $\phi$ reheats the Universe to a low reheating
temperature $\trh$, which can be smaller than $\tfo$ without
spoiling primordial nucleosynthesis \cite{hannestad} (5 MeV $\lesssim
\trh \lesssim \tfo $). Such scalar fields are  common 
in superstring models where they appear as moduli fields. The decay of
$\phi$ into radiation increases the entropy, diluting the
neutralino number density. The decay of $\phi$ into supersymmetric
particles, which eventually decay into neutralinos, increases the neutralino
number density. We denote by $b$ the net number of neutralinos
 produced on average per $\phi$ decay.

Both thermal and non thermal production mechanisms have been discussed in the literature
\cite{Moroi-Randall,kamionkowski-turner,Moroi,Chung,Giudice,Drees,Khalil,Fornengo,Pallis}. Several
supersymmetric models with particular combinations of $\trh$, $b$, and
$m_\phi$ have been
studied~\cite{Moroi-Randall,Moroi,Chung,Giudice,Drees,Khalil,Fornengo,Pallis}.
Ref.~\cite{Moroi} assumed that one neutralino is produced per $\phi$
decay ($b \simeq1$), and concluded that in the MSSM the non-thermal
neutralino production leads to an overabundance of relic neutralinos
by several orders of magnitude. Also the thermalization of
$\chi$'s produced in decays was discussed in Ref.~\cite{Moroi}.  Refs.~\cite{Chung, Giudice, Fornengo}
studied the thermal production of LSPs during the decay of a scalar
field with $b=0$, and mentioned the possibility of $b \ne 0$ but did
not study it.  Ref.~\cite{Moroi-Randall} considered purely non-thermal production.
 Ref.~\cite{Pallis} is closest to our work, in that both
thermal and non-thermal production were considered, but the general
strategy to rescue models with too low or too high standard relic
density remained, in our opinion, unclear.

In spite of all the above work, no coherent
overview of the issue, which is remarkably 
easy to understand, has to our knowledge
been laid down.  This is what we intend to
provide in this letter.

Let  $\Omega_{\rm std}$ be the density that neutralinos in a particular
model would have with the usually assumed cosmology. An appropriate
combination of the following two parameters can  bring the
relic neutralino density to the desired value $\Omega_{\rm cdm}$: the
ratio $b/m_\phi$ giving the average number of LSPs produced per unit of energy
released in each $\phi$ decay and the reheating temperature $\trh$ (which
must be $<\tfo$). We assume that the oscillating $\phi$ field
dominates the energy density of the Universe at early times and that
at $\trh$ the Universe becomes dominated by the radiation produced in $\phi$ decays.
 We follow the usual choice of the parameter $\trh$ as the 
temperature the Universe would attain under the assumption that the
$\phi$ decay and subsequent thermalization are instantaneous
\begin{equation} \Gamma_\phi = H_{\rm decay} =
\left(\frac{8\pi}{3}\right) \rho_R = \sqrt{\frac{8}{90} \pi^3g_\star}~
\frac{\trh^2}{M_P}.
\label{TRH-def}
\end{equation}
$\Gamma_\phi$ is the decay width of the $\phi$ field,
\begin{equation} \Gamma_\phi \simeq \frac{m_\phi^3}{\Lambda_{\rm
eff}^2}.
\label{gamma}
\end{equation}
If $\phi$ has non-suppressed gravitational couplings, as is usually
the case for moduli fields, the effective energy scale $\Lambda_{\rm
  eff} \simeq M_P$. In models with intermediate scales, $\Lambda_{\rm
  eff}$ could be smaller \cite{Khalil}.  Thus, with $g_\star \simeq
10$
\begin{equation} \trh \simeq 10~{\rm MeV}\left(\frac{m_\phi}{\rm
100~TeV}\right)^{3/2} \left(\frac{M_P}{\Lambda_{\rm eff}}\right).
\label{TRH}
\end{equation}
In order not to disrupt the predictions of Big Bang nucleosynthesis,
$\trh \gtrsim 5$~MeV \cite{hannestad}. Thus
\begin{equation}
m_\phi \gtrsim 100~{\rm TeV} \left( \frac{\Lambda_{\rm eff}}{M_P} \right)^{2/3}.
\end{equation}

The number $b$ of neutralinos produced per $\phi$-decay is highly
model-dependent.  It is determined by the physics of the hidden sector,
 by the mechanism of supersymmetry breaking, and in superstring-inspired
  models by the compactification mechanism. 

The coupling of the $\phi$ to the gravitino arises from the 
term $e^{K/2} \overline{\psi}_\mu \sigma^{\mu\nu} \psi_\nu$,
 where $K$ is the K\"ahler potential. If $m_\phi$ is larger than twice the gravitino mass $m_{3/2}$, the
decay mode $\phi\to \psi_{3/2}\psi_{3/2}$ of the moduli field into two
gravitinos is present with branching ratio of order 0.01 (see
Refs.~\cite{endo}, which correct previous claims~\cite{kohri} that
this branching would be chirally suppressed by a factor
$(m_{3/2}/m_\phi)^2 $). Gravitinos must then decay rapidly not to
disrupt nucleosynthesis (so $m_{3/2}\gtrsim 100 $ TeV), and they
produce comparable amounts of normal particles and their
supersymmetric partners. If $m_\phi
\gg m_{3/2}$, the gravitino decays 
 during the radiation dominated epoch after the decay of the
$\phi$ field (here we do not address this case and we focus on neutralino
production during $\phi$ domination). When $m_\phi$ and $m_{3/2}$ are
of the same order of magnitude, we can consider the gravitino decay as
part of the $\phi$ decay, since they happen almost simultaneously. In
this case, depending on how important the direct decay of $\phi$ into
supersymmetric particles other than the $\psi_{3/2}$ is, $b$ can 
typically be $0.01$--1, but not smaller. 

If instead $m_\phi < 2m_{3/2}$ more possibilities open up.  The yield
per $\phi$ decay $b$ can still be of order one but it can also be much
smaller.

Supergravity models with chiral superfields $\Phi_I$ are 
specified in terms of the K\"ahler potential $K(\Phi_I,\overline\Phi_I)$,
 the superpotential $W(\Phi_I)$, and the gauge kinetic function $f_{\alpha\beta}(\Phi_I)$. 
Specific relations between the $\phi$ mass $m_\phi$, the
 gravitino mass $m_{3/2}$, and the gaugino mass $m_{1/2}$ arise
  as a consequence of the relations $m_{3/2} = \langle e^{K/2} W \rangle$, 
  $ m_{1/2} = \langle F^J \partial_J \ln \Re f \rangle$, and 
  $m_\phi = \langle \partial^2 V/\partial \phi^2 \rangle$. With appropriate choices
   of $K$, $W$, and $f$, the  hierarchy $m_{3/2} \gtrsim m_\phi \gg m_{1/2}$ may
    be achieved. Here $V=K_{I\overline J} F^I {\overline F}^{\overline J} - 3 e^K |W|^2 + \tfrac{1}{2} (\Re f)^{-1}_{\alpha\beta} D^\alpha D^\beta$ is the scalar potential, $F^I = - e^{K/2} K^{I\overline J} D_{\overline J} {\overline W}$ is the F-term of the chiral superfield $\Phi_I$, $D^\alpha = T^{\alpha IJ} \phi_j D_IW/W$ is the D-term of the vector superfield, $D_I{\cal F} = \partial_I {\cal F} + {\cal F} \partial_I K$ is the K\"ahler covariant derivative, $\partial_I {\cal F} = \partial {\cal F}/\partial \Phi_I$, and square brackets denote vacuum values. 
\begin{figure}
\includegraphics[width=0.45\textwidth]{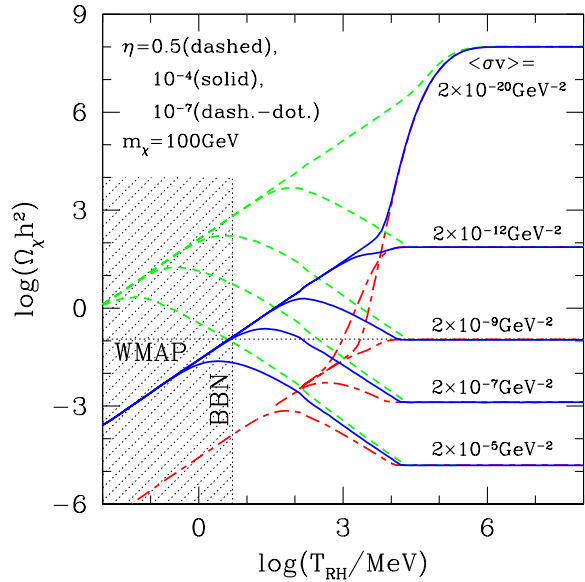}
\vspace{-10pt}
\caption{Neutralino density $\Omega_\chi h^2$ as a function of the
  reheating temperature $\trh$ for illustrative values of the ratio
  $\eta=b (100$TeV$/m_\phi)$ (the number of neutralinos produced per unit of energy
  released in $\phi$ decay, in units of (100 TeV)$^{-1}$). 
   The curves are insensitive to the initial
  conditions, i.e.\ to the value $H_I$ of the Hubble parameter at the
  beginning of the $\phi$ dominated epoch.}
\vspace{-10pt}
\end{figure}
As far as  $b$ is concerned, one finds, for example, $b\simeq
O(1)$ when the main $\phi$ decay mode is through a 
coupling of the type $ h\phi\psi^{2}$ with a
chiral matter supermultiplet $\psi$ in the superpotential $W$. This leads to comparable decay rates
of $\phi$ into the scalar and fermionic components of $\psi$ (which
are supersymmetric partners). 
On the other hand, it is possible that
the $\phi$ field decays mostly into Higgs fields, or gauge fields
(W's, Z's, photons, gluinos). In this case $b$ can
be very small $10^{-2},~10^{-4},~10^{-6}$ etc.~\cite{Moroi-Randall, Drees}.
For example, the coupling of $\phi$  to the gauge bosons arises from the
 term $ \Re f_{\alpha\beta} F^\alpha_{\mu\nu} F^{\mu\nu\beta}$ 
 and with non-minimal kinetic terms $f_{\alpha\beta}$ may contain
$\phi$. The $\phi$ decay width into gauge bosons is then
$\Gamma_g \sim \lambda_g m_\phi^3/M_P^2$ with
$\lambda_g = \frac{\partial}{\partial\phi} \ln \Re f$, while that into 
gauginos is $\Gamma_{\tilde g} \sim \lambda_{\tilde g}^2 m_\phi/M_P^2$ with
$\lambda_{\tilde g} = m_{1/2} \frac{ \partial}{\partial\phi} \ln (F^\phi \frac{\partial}{\partial\phi} \Re f)$.
Thus in principle the gaugino coupling may be suppressed relative
 to the coupling to gauge bosons. 

 Neutralinos are in kinetic equilibrium with the radiation when their scattering rate off
relativistic particles is faster than the Hubble expansion rate,
$\Gamma^{\rm scatt} \gtrsim H$. During the epoch in which the Universe
is dominated by the decaying $\phi$ field, $H$ is proportional to
$T^4$~ \cite{McDonald}. In fact, in the evolution equation for the radiation energy density (Eq. (5) below with
$p=\rho/3$) substitute $\rho \simeq T^4$ and 
$\rho_\phi \simeq M_P^2 H^2$. Then use $H\sim t^{-1}$, write $T \propto t^{\alpha}$,
match the powers of $t$ in all terms, and determine  $\alpha=-(1/4)$.
Hence, during the oscillating
$\phi$ dominated epoch $H \propto t^{-1} \propto T^4$
 (and $\rho_{\phi} \propto H^2  \propto T^8$).
Since  $H$ 
 equals $\trh^2/M_P$ at $T=\trh$, it is $H \simeq T^4/(\trh^2
M_P)$. Since $\Gamma^{\rm scatt} \simeq n_\gamma \sigma_{\rm scatt}
\simeq T^3 \sigma_{\rm scatt}$, we see that kinetic equilibrium is
maintained at low temperatures, $T \lesssim \trh^2 M_P \sigma_{\rm
  scatt}$. This is contrary to the usual radiation-dominated scenario
in which the temperature has to be large enough for kinetic
equilibrium to be maintained. The origin of this difference is the
strong $T^4$ dependence of $H$ on the radiation temperature. Using the known
relation $\Omega_{\rm std} h^2 \simeq 10^{-10} {\rm
  ~GeV^{-2}}/\sigmav$, and taking $\sigma_{\rm scatt}$
of the same order of magnitude as the annihilation cross section $\langle\sigma
v\rangle$, gives $T \lesssim (10^6 {\rm ~MeV~}/\Omega_{\rm std} h^2)
(\trh/{\rm MeV})^2$. For the following we need to assume kinetic
equilibrium before neutralino production ceases: the right hand
side is larger than $\trh$ for $\Omega_{\rm std}h^2 \lesssim 10^6
(\trh/{\rm MeV})$, so we will safely assume that kinetic equilibrium is
reached.

In kinetic equilibrium the equations which describe the
evolution of the Universe are:
\begin{eqnarray} 
\label{eq:evol1}
\dot\rho &=& -3H(\rho+p)+\Gamma_\phi \rho_\phi \\ \dot n_\chi &=&
-3Hn_\chi-{\sigmav}
\left(n_\chi^2-n^2_{\chi,\rm eq}\right) + \frac{b}{m_\phi} \, \Gamma_\phi \rho_\phi \\ \dot
\rho_\phi &=&-3H \rho_\phi -\Gamma_\phi \rho_\phi \\ H^2 &=& \frac{8\pi}{3M_P^2}
(\rho + \rho_\phi).
\label{eq:evol4}
\end{eqnarray}
Notice that the equations, and thus the results, depend only on the
combination $b/m_\phi$ and not on $b$ and $m_\phi$ separately.
 In Eqs.~(\ref{eq:evol1}-\ref{eq:evol4}), a dot indicates a time
derivative, $\rho_\phi$ is the energy density in the $\phi$ field,
which is assumed to behave like non-relativistic matter; $\rho$ and
$p$ are the total energy density and pressure of matter and radiation
at temperature $T$, which are assumed to be in kinetic but not
necessarily chemical equilibrium; $n_\chi$ is the number density of
LSPs, and $n_{\chi,\rm eq}$ is its value in chemical equilibrium;
finally, $H=\dot a/a$ is the Hubble parameter, with $a$ the scale
factor. For convenience in the numerical calculations, we used as
independent variable $\ln a$ and as dependent variables $Y = n_\chi/
s$, $Y_\phi =\rho_\phi/(m_\phi s)$ and $T$, where $s=(\rho+p-m_\chi
n_\chi)/T$ is the entropy density of the matter and radiation. We also
used the first principle of thermodynamics in the form $d(\rho
a^3)+d(\rho_\phi a^3)+pda^3=Td(s a^3)$ to rewrite
Eq.~(\ref{eq:evol1}) as
\begin{equation} \dot s = - 3 H s + \frac{\Gamma_\phi \rho_\phi}{T}.
\end{equation}

Initial conditions are specified through the value $H_I$ of the Hubble
parameter at the beginning of the $\phi$ dominated epoch. This amounts
to giving the initial energy density $\rho_{\phi,I}$ in the $\phi$
field, or equivalently the maximum temperature of the radiation
$T_{\rm MAX}$. Indeed, one has $H_I \simeq \rho_{\phi,I}^{1/2}/M_P
\simeq T_{\rm MAX}^4/(\trh^2 M_P)$. The latter relation can be derived from
$\rho_\phi \simeq T^8/\trh^4$ and the consideration that the maximum
energy in the radiation equals the initial (maximum) energy
$\rho_{\phi,I}$. 

If the neutralino reaches chemical equilibrium, it is clear that its
final density does not depend on the initial conditions. 
An approximate condition for reaching chemical equilibrium is \cite{Giudice}
$\sigmav \gtrsim 10^{-9} {\rm GeV}^{-2}$ $(m_\chi/100{\rm
  GeV}) (\trh/{\rm MeV})^{-2}$. Even without reaching chemical equilibrium, the
neutralino density is insensitive to the initial conditions provided the
maximum temperature of the radiation $T_{\rm MAX} \gtrsim
m_\chi$ \cite{Giudice}.

In Fig.~1 we show how the neutralino density $\Omega_{\chi} h^2$ depends on
 $\trh$ for illustrative values of the
parameter $\eta = b (100{\rm TeV}/m_\phi)$, both for neutralinos which are underdense and
which are overdense in usual cosmologies.

The behavior of the relic density as a function of $\trh$ is easy to
understand physically.
If $\sigmav$ is large enough so that chemical equilibrium is achieved,
the usual thermal production scenario occurs for $\trh > \tfo$.
Neutralino annihilation compensates thermal production
 until the latter ceases to be effective at $T=\tfo$. The LSP density is then
determined by the condition $\Gamma^{\rm ann} = n \sigmav \simeq H$ at
$T=\tfo$. Using $H \simeq T^2/M_P$, as appropriate for a radiation
dominated Universe, this leads to $ Y_0 \simeq Y_{\rm f.o.} \simeq
(n/s)_{\rm f.o.} \simeq (H/s\sigmav)_{\rm f.o.} \simeq 1/(\tfo M_P
\sigmav)$. This gives the usual result $\Omega_{\rm std} \simeq (m
s_0/\rho_c) Y_0 \simeq (m s_0)/(\rho_c \tfo M_P \sigmav) \simeq 2
\times 10^{-10} {\rm GeV}^{-2}/\sigmav $. Here we used $s_0/\rho_c =
2.8 \times 10^8 {\rm GeV}^{-1}$. When $\sigmav$ is too small for
chemical equilibrium to be achieved, the usual equation does not
hold. Notice that this is the case for the smallest $\sigmav$
in Fig.~1.

For $\trh<\tfo$, there are four different ways in which the density
$\Omega h^2$ depends on $\trh$:

(1) Thermal production without
chemical equilibrium. In this case  $\Omega_{\chi} \propto \trh^7$ (e.g.  steepest
part of the $\eta=10^{-7}$, $\sigmav=2\times 10^{-20}$ GeV$^{-2}$ line).
The relic density was estimated
in Ref.~\cite{Chung}:
\begin{equation} \frac{ \Omega_\chi}{\Omega_{\rm cdm}} \simeq 
\frac{\sigmav}{10^{-16}~{\rm GeV}^{-2}}
\left(\frac{100{\rm GeV}}{m_\chi}\right)^{5}
\left(\frac{\trh}{{\rm GeV}}\right)^7
\left(\frac{10}{g_\star}\right)^{3/2}.
\end{equation}
This matches our numerical calculation for $\sigmav = 2\times 10^{-20}$ GeV$^{-2}$.

(2) Thermal production with chemical equilibrium. In this case
 $\Omega_{\chi} \propto
\trh^4$ (e.g. steepest part of the $\eta=10^{-7}$,
$\sigmav=2\times 10^{-12}$ GeV$^{-2}$ line).
The neutralino freezes out while the universe is
dominated by the $\phi$ field. Its freeze-out density is larger than
usual, but it is diluted by entropy production from
$\phi$ decays. The new freeze-out temperature $\tfo^{\rm NEW}$ is determined
by solving $n \sigmav \simeq H$ at $T=$ $\tfo^{\rm NEW}$. Using the relations
between $H$, $a$, and $T$ in the decaying-$\phi$ dominated Universe,
one finds \cite{McDonald,Giudice} $\Omega_\chi \simeq$
$\trh^3\tfo(\tfo^{\rm NEW})^{-4} \Omega_{\rm std}$. Our numerical results
indicate a slope closer to $\trh^4$, perhaps due to the change in
$\tfo$.

 (3) Non-thermal production
without chemical equilibrium. Here $\Omega_{\chi} \propto \trh$ (e.g. 
leftmost part of each line).
Non-thermal production is not compensated by annihilation. 
The production of neutralinos is purely
non-thermal and the relic density depends on $\eta$. It can be estimated
analytically as follows. For each superpartner produced, at least one LSP will
remain at the end of a chain of decays (due to $R$-parity
conservation), and thus $n_{\chi} \simeq bn_\phi$. Here $n_\phi =
\rho_\phi/m_\phi$.  At the time of $\phi$-decay
$ \rho_{\chi} \simeq m_\chi b \rho_\phi/m_\phi \simeq \trh^4$, and the
entropy  is $s \simeq \trh^3$. Hence $
\rho_\phi/s\simeq \trh$ and $Y_0=Y_{\rm decay}\simeq b \trh/m_\phi$. It follows that,
\begin{equation} 
\frac{\Omega_{\chi}}{\Omega_{\rm cdm}} \simeq 2 \times 10^3 \eta \left(\frac{m_\chi}{\rm
100~GeV}\right) \left(\frac{\trh}{\rm MeV}\right)
\label{nonthermal}
\end{equation}

(4) Non-thermal production with
chemical equilibrium. In this case $\Omega \propto \trh^{-1}$ (e.g.  central
part of the $\eta=0.5$, $\sigmav=2\times 10^{-5}$ GeV$^{-2}$ line).
Here annihilation compensates for the
non-thermal production of neutralinos until the non-thermal production ceases at
$T=\trh$. The condition for determining the relic density is
$\Gamma^{\rm ann} \simeq \Gamma_{\phi}$ at $T=\trh$. This leads to $ Y_0
\simeq Y_{\rm RH} \simeq \Gamma_\phi/(s_{\rm RH} \langle\sigma v
\rangle) \simeq 1/(\trh M_P \sigmav)$. From here it follows that 
\begin{equation}
\Omega_\chi \simeq (\tfo/\trh) \Omega_{\rm std}. 
\label{eq:case4}
\end{equation}

With the help of Fig.~1 and formulas (\ref{nonthermal}) and
(\ref{eq:case4}), we can separate the different ranges of $\eta$.
Notice that in Fig.~1 $m_\chi = 100 $ GeV, thus 
$\tfo \simeq m_\chi/20 \simeq 5$ GeV.  For $\trh >
\tfo$, the standard production mechanism is recovered, thus $\Omega_\chi =
\Omega_{\rm std}$ (indicated by the horizontal lines on the right of
the Figure). If $\trh < \tfo$, the value of $\Omega h^2$ depends on
$\eta$.  Overdense neutralinos, i.e.\ those
with $\Omega_{\rm std} > \Omega_{\rm cdm}$ (above the dotted line labelled
WMAP), require values of  $\eta \lesssim 10^{-4} (100{\rm GeV}/m_\chi)$
to bring their density to $\Omega_\chi=\Omega_{\rm cdm}$. This bound is
derived from the BBN condition $\trh \gtrsim 5$ MeV by taking
$\Omega_\chi = \Omega_{\rm cdm}$ in eq.~(\ref{nonthermal}). The
condition $\trh \le \tfo$ in Eq.~(\ref{nonthermal}) shows that for a
solution $\Omega_\chi = \Omega_{\rm cdm}$ with $\eta \lesssim 10^{-7}
(100 {\rm GeV}/m_\chi)^2$ the production must be thermal with entropy
dilution (case (1) or (2)), and $\trh$ must be close to $\tfo$. Notice that in between
the two values of $\eta$ just mentioned the production is purely
non-thermal (no chemical equilibrium, namely case (3)). For all overdense
neutralinos, given one value of $\eta\lesssim 10^{-4} (100{\rm
  GeV}/m_\chi)$ there is only one value of $\trh$ for which
$\Omega_\chi = \Omega_{\rm cdm}$.

Underdense neutralinos, i.e.\ those with $\Omega_{\rm std} <
\Omega_{\rm cdm}$, can have zero, one, or two solutions $\Omega_\chi=\Omega_{\rm cdm}$. There is no
solution if $\Omega_{\rm std}$ is too low. Neutralinos with
$\Omega_{\rm std} \lesssim 10^{-5} (100 {\rm GeV}/m_\chi)$ cannot be
brought to $\Omega_{\rm cdm}$, independently of $\eta$. This can be seen by imposing the
condition $\trh > 5$ MeV and $ \Omega_\chi = \Omega_{\rm cdm}$ in
Eq.~({\ref{eq:case4}). Neutralinos with $\Omega_{\rm cdm} \gtrsim \Omega_{\rm std} \gtrsim 10^{-5} (100 {\rm GeV}/m_\chi)$ cannot be brought to $\Omega_{\rm cdm}$ either if $\eta \lesssim 10^{-7} (100 {\rm GeV}/m_\chi)^2  (\Omega_{\rm cdm}/\Omega_{\rm std})$. This happens when the non-thermal production is insufficient to increase the density up to $\Omega_{\rm cdm}$, such as  for the $\eta=10^{-7}$, $\sigmav=2\times 10^{-7}$ GeV$^{-2}$ line in Fig.~1. Again using Eqs. (\ref{nonthermal}) and
(\ref{eq:case4}) and Fig.~1, one can see that
for the same range of densities $\Omega_{\rm cdm} \gtrsim\Omega_{\rm std} \gtrsim 10^{-5} (100 {\rm GeV}/m_\chi)$ there are two solutions for $ 10^{-7}(100 {\rm GeV}/m_\chi)^2 (\Omega_{\rm cdm}/\Omega_{\rm std}) \lesssim \eta \lesssim 10^{-4}(100 {\rm GeV}/m_\chi)$ and a single solution for larger values of $\eta$.
  The two solutions have
  different values of $\trh$ and are both non-thermal one belonging to
  case (3) and the other to case~(4). The single solution belongs to case~(4). 

In conclusion the neutralino density can be that of cold dark matter provided
  $\Omega_{\rm std} \gtrsim 10^{-5} (100 {\rm GeV}/m_\chi)$ and the high energy theory
  accomodates the combinations of values of $b/m_\phi$ and
  $\trh$  identified in the previous two paragraphs. These conditions may place constraints on high-energy models derived from superstring theories.
  
This work was supported in part by the US Department of Energy Grant
DE-FG03-91ER40662, Task C and NASA grants NAG5-13399  and ATP03-0000-0057
at UCLA, and NFS
grant PHY-0456825 at the University of Utah. We thank A. Masiero,
S. Nussinov, A. Soldatenko, and C. Yaguna for helpful discussions.

\end{document}